\documentclass{IEEEtran}
\usepackage{cite}
\usepackage{amsmath,amssymb,amsfonts}
\usepackage{algorithmic}
\usepackage{graphicx}
\usepackage{textcomp}
\usepackage{lipsum,mathtools,cuted}
\usepackage{xcolor}
\def\BibTeX{{\rm B\kern-.05em{\sc i\kern-.025em b}\kern-.08em
    T\kern-.1667em\lower.7ex\hbox{E}\kern-.125emX}}
    
\def\e{\begin{equation}}
\def\f{\end{equation}}
\def\_#1{{\bf #1}}

\def\E{\varepsilon}

\def\.{\cdot}

\def\=#1{\overline{\overline #1}}

\def\l#1{\label{eq:#1}}
\def\r#1{(\ref{eq:#1})}
\def\@#1{_{\rm #1}}

\begin{document}
\title{Metasurface retroreflector for controlling parasitic surface waves on bodies of 5G mobile phones}

\author{F.S.~Cuesta, V.A.~Lenets, A.~D{\'i}az-Rubio, A.~Khripkov, and S.A.~Tretyakov

\thanks{(Corresponding author: Francisco S. Cuesta)}
\thanks{F.S.~Cuesta, A.~D{\'i}az-Rubio, and S.A.~Tretyakov 
are with the Department of Electronics and Nanoengineering, Aalto University, 
FI-00079 Aalto, Finland 
(e-mail: francisco.cuestasoto@aalto.fi;
 ana.diazrubio@aalto.fi; sergei.tretyakov@aalto.fi).}
\thanks{V.A.~Lenets was with the Faculty of Physics and Engineering, ITMO University, Russia.}
\thanks{A.~Khripkov is with the Terminal Antenna and RF Laboratory, Huawei Technologies, Helsinki, Finland. (e-mail: alexander.khripkov@huawei.com).}
}

\maketitle

\begin{abstract}

The smartphone, as a media device, should offer a large-size screen and fast data rates in a well-packed device. Emergent 5G technologies based on millimeter waves need  efficiently radiating antennas integrated into devices whose bodies (especially glass screens) behave as  waveguides in the antenna frequency bands. Using the concept of retroreflective surfaces, this work proposes a compact metasurface located below the smartphone glass screen which reduces surface-wave excitation and propagation, and significantly improves the radiation pattern and efficiency of  millimeter-wave antennas integrated into mobile terminals.

\end{abstract}

\begin{IEEEkeywords}
Millimeter wave, mobile phone antenna, 5G,  surface waves, retroreflector, metasurface

\end{IEEEkeywords}

\section{Introduction}
\IEEEPARstart{F}{rom} watching a video to writing a scientific article, the smartphone has become the 21st century Swiss Army knife: compact yet versatile.  As a streaming device, the smartphone requires high-speed communication solutions. It is expected that 5G technologies in the  millimeter-wave (mmWave) range \mbox{$f = 20 \-- 40$ GHz} \mbox{($\lambda = 7.5 \-- 15$ mm)}  \cite{Viikari_2019, Yong_2019, He_2020, Ikram_2020} could grant transfer rates in the order of Gbps \cite{Albreem_2015, Qi_2016, Lu_2017}. However, practical  implementation becomes troublesome as the wavelength is  smaller than the device dimensions.

\begin{figure}[!h]
\centerline{\includegraphics[width=0.8\columnwidth]{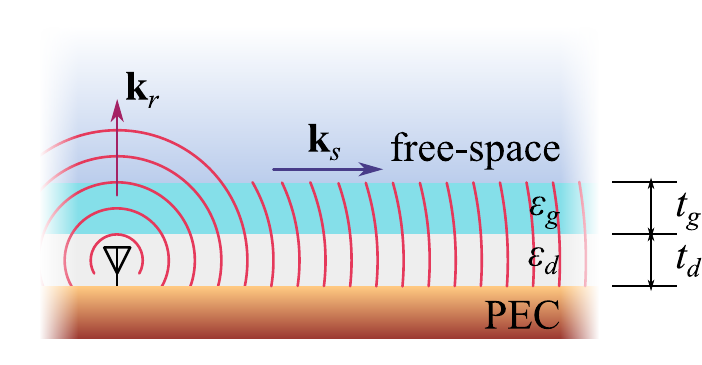}}
\caption{Schematic representation of the problem, where the smartphone is simplified as a composition of a few material layers. The chassis, modelled as a PEC layer, is covered by a dielectric layer  with thickness $t_d$ and relative permittivity $\E_d$. A glass layer, with thickness $t_g$ and permittivity $\E_g$ is placed over the dielectric layer. An antenna placed inside the structure will not only radiate waves into surrounding space, with the propagation constant $\_k\@{r}$, but also will excite surface waves propagating with $\_k\@{s}$ along the layers.}
\label{schem}
\end{figure}

In order to achieve a wide-beam full-sphere coverage and minimize user blockage, one of the following designs of a mobile device can be used: three end-fire antenna array modules on the top, left and right sides, respectively; or two antenna arrays at the front and back sides, respectively. It was shown in  \cite{Raghavan_2019,Hazmi_2019} that both designs have comparable performance in terms of beamforming antenna gain patterns and link budgets, and robust performance with hand blockage.
Implementation of end-fire antennas is limiting the display size and display edge curvature of the mobile device. Preferable edge areas with statistically less impact of the hand blockage are the same for the end-fire mmWave array modules and for sub-6 GHz antennas. Positioning of mmWave end-fire antenna array modules impose limitations on the space available for sub-6 GHz antennas. Therefore, display-side and backside mmWave antenna array modules are beneficial from the full-screen design and multiband wireless connectivity considerations.

\begin{figure}[!h]
\centerline{\includegraphics[width=\columnwidth]{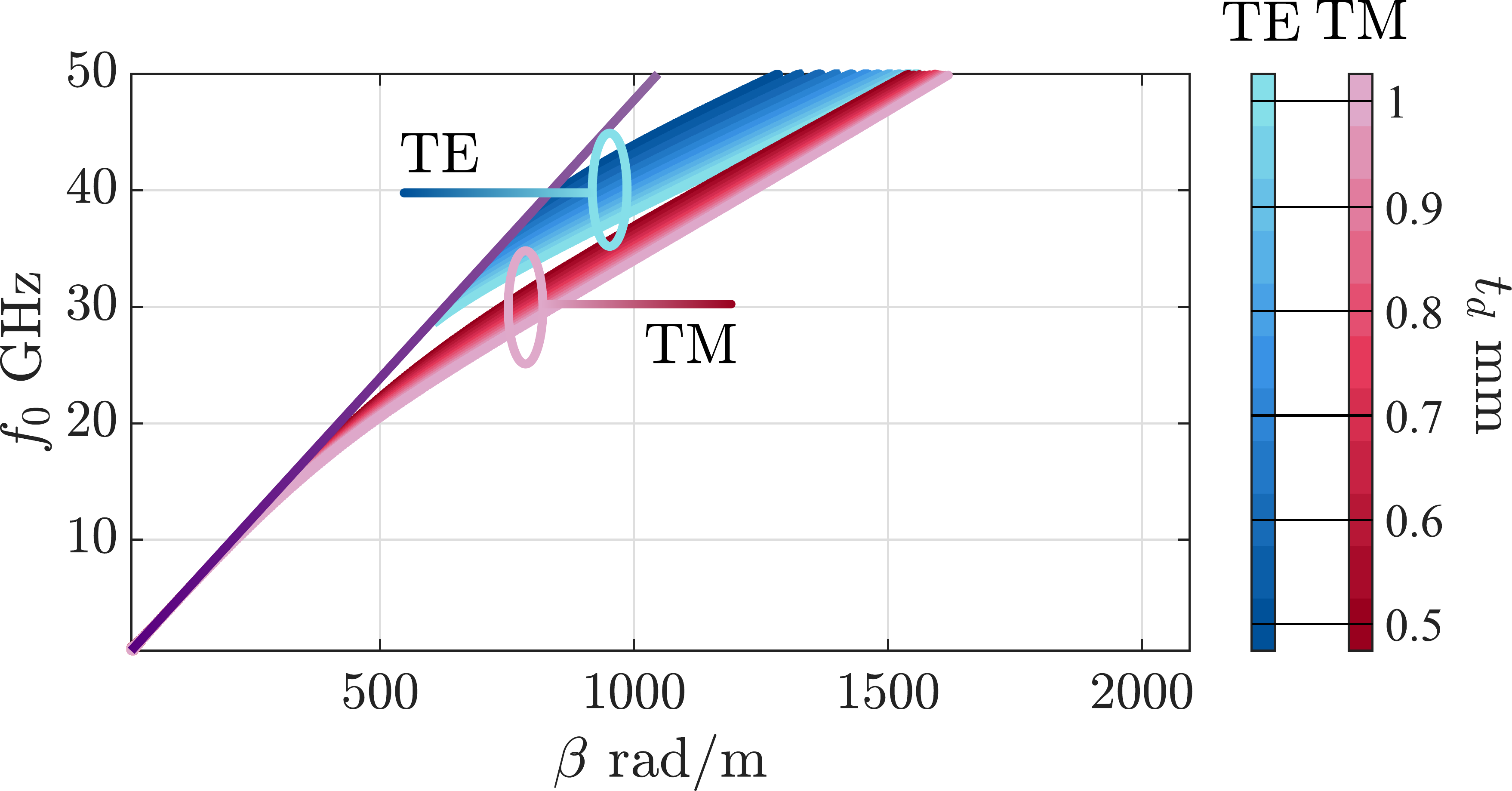}}
\caption{Dispersion diagram for different thicknesses $t_d$, for dielectric permittivity $\varepsilon_{d} = 2.7$, glass permittivity $\varepsilon_{d} = 5.5$ and glass thickness \mbox{$t_g=0.5$ mm} for TE and TM modes. 
These results were obtained numerically using computational software MATLAB.}
\label{fig:eig_theor}
\end{figure}

In conceptual studies of device-integrated  millimeter-wave antennas, a smartphone can be considered as a multilayered structure consisting of a glass layer, a dielectric layer (which combines the tactile panel and the screen display) with some effective permittivity, over a metallic chassis. A 5G antenna located under the glass, as shown in \mbox{Fig.~\ref{schem}}, would excite surface waves along the glass and dielectric layers, as in a dielectric-slab waveguide. 
The dispersion properties of surface waves, both TE- and TM-polarized modes, can be deduced analytically using the vector transmission-line method \cite{Tretyakov_2003}. The dispersion equations for TE- and TM-modes read, respectively,
\begin{subequations}\label{eq:eig_mod}
\begin{align}
	\frac{j}{\gamma_{g}} \frac{\gamma_{g} \tan \left(t_{d} \gamma_{d}\right)+\gamma_{d} \tan \left(t_{g} \gamma_{g}\right)}{\gamma_{d}-\gamma_{g} \tan \left(t_{g} \gamma_{g}\right) \tan \left(t_{d} \gamma_{d}\right)}&=\frac{1}{\sqrt{k_{0}^{2}-\beta^{2}}}, \label{eq:TE_eig}\\
	\frac{j \gamma_{g} \varepsilon_{g} \gamma_{d} \tan \left(t_{d} \gamma_{d}\right)+\varepsilon_{d} \gamma_{g} \tan \left(t_{g} \gamma_{g}\right)}{\varepsilon_{d} \gamma_{g}-\varepsilon_{g} \gamma_{d} \tan \left(t_{g} \gamma_{g}\right) \tan \left(t_{d} \gamma_{d}\right)}&=\sqrt{k_{0}^{2}-\beta^{2}}, \label{eq:TM_eig}\\
	\gamma_{g} = \sqrt{\varepsilon_{g} k_{0}^{2}-\beta^{2}},& \label{eq:gamma_g}\\
	\gamma_{d} = \sqrt{\varepsilon_{d} k_{0}^{2}-\beta^{2}},& \label{eq:gamma_d}	
\end{align}
\end{subequations}
where $t_g$ and $\varepsilon_g$ are the glass thickness and relative permittivity, respectively, while $t_d$ and $\varepsilon_d$ are the respective parameters for the dielectric layer. The dispersion diagram of \mbox{Fig.~\ref{fig:eig_theor}} shows that, with the typical dimensions of commercial smartphones, an antenna behind the glass can excite TM waves at the frequencies as low as 20~GHz, while TE modes propagate starting from approximately 30~GHz. 

Surface wave excitation distorts the desired radiation pattern, compromising scattering along the multi-layer plane.  Moreover, the presence of the glass and dielectric layers  produces mismatch between the 5G antenna and the free space, which decreases the overall efficiency. 
In practical terms, the waveguide supported by the reinforced glass covers at the front and back sides of the mobile device results in additional dissipative loss of 1--3~dB \cite{Syrytsin_2017,Qualcomm_2018}. 
Mobile device antennas utilize various techniques for controlling current distribution on the ground plane, such as a  parasitic inverted L-antenna (ILA) element located at the ground edge \cite{Islam_2011} or wavetraps at the antenna connection to the ground plane \cite{Ying_2017}. 

It shall be noted that typically surface wave suppression is used  for  improvement of isolation between adjacent antennas, and currents on the ground plane edges are suppressed by electromagnetic bandgap (EBG) unit cells \cite{Lee_2019}.
From the analysis of different approaches, found in the literature \cite{Kildal_1990, Sievenpiper_1999, Wu_2007, Mulenga_2010, Kumar_2015, Oraizi_2018}, it is clear that  volumetric, 3D unit cells cannot  be easily implemented inside  dielectric layers; while a flat structure, more friendly in terms of implementation, can be capable of  directing  electromagnetic waves away from the surface. It is important that these planar structures can be designed to remain practically invisible to the human eye \cite{Park_2019,Stanley_2018}, as the solution must not affect display visibility. Moreover, a compact structure will reduce potential electromagnetic compatibility issues. 

We expect that recently conceptualized metasurface retroreflectors \cite{Asadchy_2017_flat,Shen_2018,Song_2018}, which are capable of reflecting an incident wave back along the illumination direction, can be a valuable solution. For this application, it is expected that a retroreflecting metasurface can work as a decoupling structure between the antenna and the layered body of the device, sending the waves created by the antenna back into space (primarily along the incidence direction and with a controllable reflection phase) and contributing to better radiation performance. In contrast to conventional stop-band solutions discussed above, the propagation of the surface wave is not suppressed. Instead, these waves are re-radiated into space, preferably along the desired direction. These retroreflective devices can be implemented designing metasurfaces which ensure the desired phase matching between the incident and reflected waves \cite{Wang_2018_asymmetry,Nagayama_2016,Arbabi_2017,Yan_2018}. 

In this paper, we present a solution based on the use of retroreflective metasurfaces, which allows us to minimize the propagation of surface waves across an all-screen device in the mmWave regime, when the antenna is placed behind the glass, close to the screen edge. The paper is organized as following: First, we introduce the retroreflector solution and  explain its operation principle, and then present the methodology and design results. In the results section, the performance of the retroreflective surface is validated using full-wave simulations.

\section{Retroreflectors for surface wave control}

\begin{figure}[!h]
\centerline{\includegraphics[width=0.6\columnwidth]{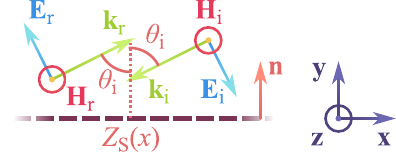}}
\caption{A  retroreflective boundary sends back a TM-polarized wave incident at the incidence angle $\theta_i$. To ensure phase matching between the incident and reflected waves, the surface impedance is made inhomogeneous along the $x$-coordinate.}
\label{rr_TM}
\end{figure}

\subsection{Surface Impedance}

A retroreflective surface can be characterized by its surface (input) impedance $Z\@s$ that relates the total (incident plus reflected) tangential electric $\_E_{\tau}$ and magnetic fields $\_H_{\tau}$ at its surface:
\begin{equation}
Z\@{s} \hat{y} \times \_H_{\tau}=\_E_{\tau}, \l{surf_imp}
\end{equation}
where  $\hat{y}$  is the unit vector normal to the surface \cite{Tretyakov_2003}. The retroreflector is designed for TM-polarized waves, because surface waves of TE polarization can propagate through the device only above 30~GHz, see  \mbox{Fig.~\ref{fig:eig_theor}}. Using the representation of a TM-retroreflector of \mbox{Fig.~\ref{rr_TM}}, the incident and reflected waves are written, respectively, as
\begin{subequations}\label{eq:inc_field}
\begin{align}
	\_E\@{i} &= \eta  H_{0} e^{j k_{0}\left( x \sin \theta\@{i}+ y \cos \theta\@{i}\right)} \left(\cos \theta\@{i} \hat{x} - \sin \theta\@{i} \hat{y} \right), \label{eq:inc_e}\\
	\_H\@{i} &= H_{0} e^{j k_{0}\left( x \sin \theta\@{i}+ y \cos \theta\@{i}\right)} \hat{z}, \label{eq:inc_m}	
\end{align}
\end{subequations}
\begin{subequations}\label{eq:rflc_field}
\begin{align}
	\_E\@{r} &= \eta  R H_{0} e^{-j k_{0}\left( x \sin \theta\@{i} + y \cos \theta\@{i}\right)} \left(-\cos \theta\@{i} \hat{x} + \sin \theta\@{i} \hat{y} \right), \label{eq:rflc_e}\\
	\_H\@{r} &= R H_{0} e^{-j k_{0}\left( x \sin \theta\@{i} + y \cos \theta\@{i}\right)} \hat{z}, \label{eq:rflc_m}	
\end{align}
\end{subequations}
where $R = |R|\exp \left[j\varphi\@{r}\right]$ is the reflection coefficient, $\theta\@{i}$ is the incidence angle, and $\eta$ is the wave impedance of the medium above the retroreflective surface. The corresponding total tangential magnetic and electric fields, required to define $Z\@{s}$, are the sums of the incident and reflected fields at the interface ($y=0$): 
\begin{subequations}\label{eq:tang_field}
\begin{align}
	\_E_{\tau} &= \eta  H_{0} \cos \theta\@{i} \left(e^{j k_{0} x \sin \theta\@{i}}  - R  e^{-j k_{0} x \sin \theta\@{i}}\right)\hat{x}, \label{eq:tang_e}\\
	\_H_{\tau} &= H_{0} \left(e^{j k_{0} x \sin \theta\@{i}}  + R  e^{-j k_{0} x \sin \theta\@{i}}\right) \hat{z}. \label{eq:tang_m}	
\end{align}
\end{subequations}
The surface impedance for a retroreflective surface reads
\begin{equation}
Z_{s}=\eta \cos \theta\@{i} \frac{1-|R| e^{j \phi}}{1+|R| e^{j \phi}},
\label{mts_imp_any_r}
\end{equation}
where $\phi=-2 k_{0} x \sin \theta\@{i}+\varphi\@{r}$ is the phase gradient required for the proper operation of the metasurface. In case of total retroreflection ($|R|=1$), the surface impedance is reduced to 
\begin{equation}
Z_{s}=-j\eta \cos \theta\@{i} \tan \left(\dfrac{\phi}{2}\right).
\label{mts_imp}
\end{equation}
These results show that the retroreflective surface is a lossless periodic structure (with the period $D=\frac{\lambda}{2 \sin \theta\@{i}}$) and strongly frequency and incidence-angle dependent. The period increases when the incidence angle decreases, and in the limit of zero angle (normal incidence), the retroreflector degenerates to a usual uniform mirror.

\subsection{Grid Impedance}
\begin{figure}[!h]
\centerline{\includegraphics[width=0.6\columnwidth]{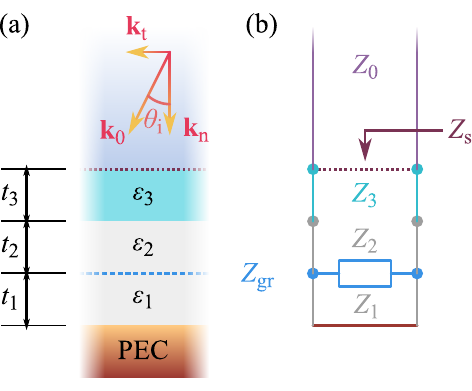}}
\caption{(a) The structure under study  can be seen as a combination of layers with different relative permittivities $\varepsilon_n$ and thicknesses $t_n$, with $n=1,2,3$ denoting the different layers. (b) The designed metasurface, located in the middle of the dielectric layer, has a grid impedance $Z\@{gr}$ capable to create a virtual retroreflector at the glass  surface.}
\label{eqv_circ}
\end{figure}

In the previous subsection, we modeled the retroreflector as an idealized  surface impedance  boundary. In practice, the retroreflector can be  realized as a multilayer structure consisting of two dielectric layers and a structured metal film placed in the middle of the second layer, as shown in \mbox{Fig. \ref{eqv_circ}(a)}. The structure  should be designed to produce a virtual retroreflector at the top glass surface, as explained in the equivalent system circuit in \mbox{Fig.~\ref{eqv_circ}(b)}, where the input impedance at the top of the glass layer is equal to the retroreflector surface impedance. The structured metal film can be viewed as a single-layer metasurface, which  can be characterized by a grid impedance $Z\@{gr}$ that defines  a discontinuity of the total tangential magnetic field at the metasurface plane. 

\begin{table*}
\begin{equation}
Z\@{gr}=\frac{Z_{1} Z_{2} \delta_{1}\left[Z_{3}\left(Z_{2} \delta_{2}+Z_{3} \delta_{3}\right)-j Z\@{s}\left(Z_{2} \delta_{2} \delta_{3}-Z_{3}\right)\right]}{j Z_{3}\left[Z_{2}\left(Z_{2} \delta_{2}+Z_{3} \delta_{3}\right)+Z_{1} \delta_{1}\left(Z_{2}-Z_{3} \delta_{2} \delta_{3}\right)\right]+Z\@{s}\left[Z_{2}\left(Z_{2} \delta_{2} \delta_{3}-Z_{3}\right)+Z_{1} \delta_{1}\left(Z_{2} \delta_{3}+Z_{3} \delta_{2}\right)\right]}
\protect
\l{sandwich_eqv}
\end{equation}
\end{table*}

Using the transmission-line theory \cite{Ulaby_2006}, we can calculate the input impedance of the multilayer system $Z\@{in}$ and equate it to the required  surface impedance of a retroreflective surface $Z\@{s}$, as shown in \mbox{Fig.~\ref{eqv_circ}}. Each dielectric layer behaves as a transmission line, with the corresponding TM characteristic impedance and equivalent propagation constant \cite{Tretyakov_2003}.
After equating the input impedance of the multilayer system with the retroreflector surface impedance, the required grid impedance can be written as in Eq.~\r{sandwich_eqv}, where \mbox{$\delta_{n}=\tan \left(t_{n} k_{0} \sqrt{\varepsilon_{n}-\sin ^{2}\theta\@{i}}\right)$} and \mbox{$Z_{n}=\frac{\eta_{0}}{\varepsilon_{n}} \sqrt{\varepsilon_{n}-\sin ^{2}\theta\@{i}}$} with $n \in[1,2,3]$ representing the three dielectric layers.

\begin{figure}[!h]
\centerline{\includegraphics[width=\columnwidth]{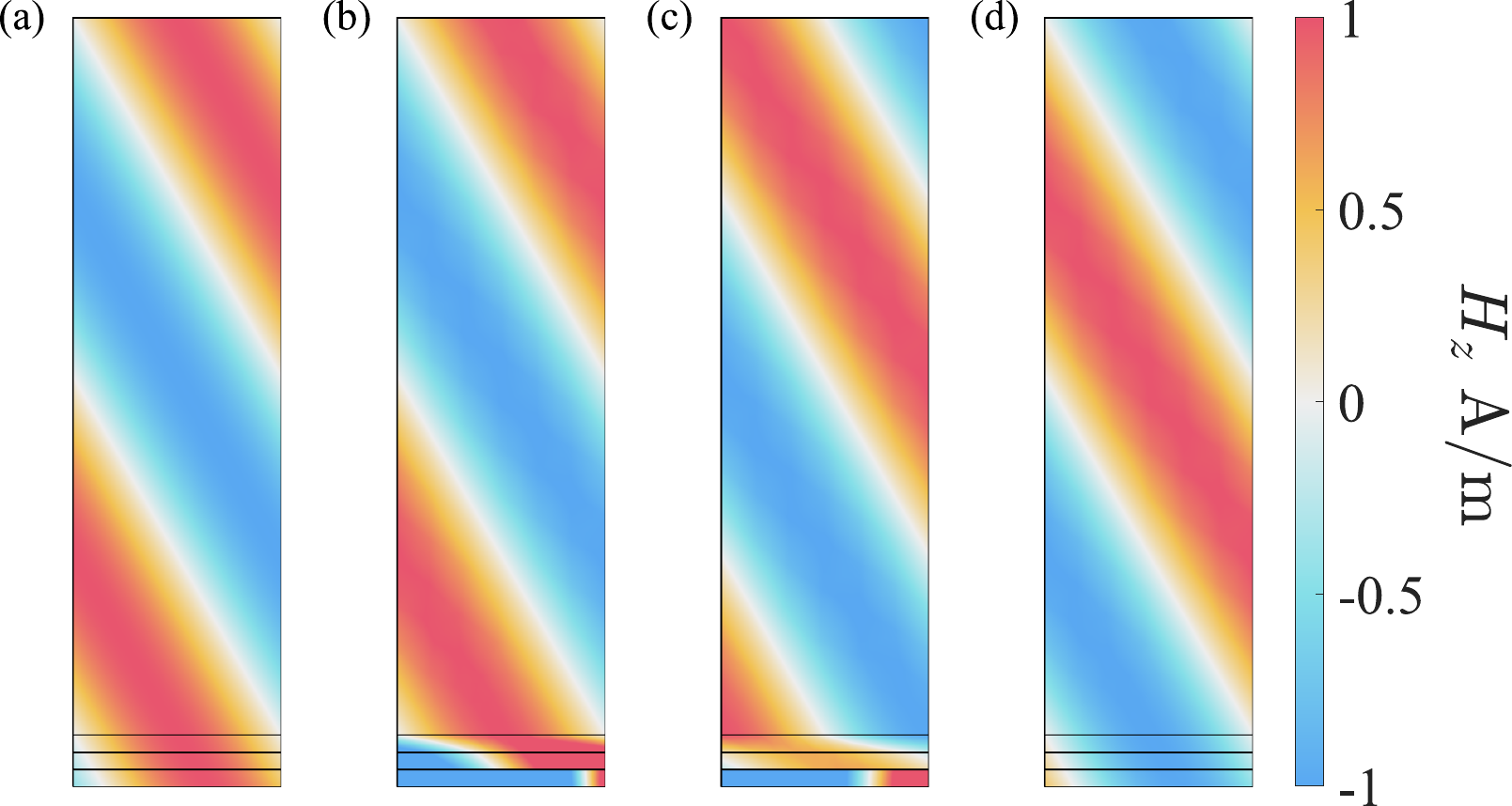}}
\caption{A retroreflective metasurface tuned for $\theta_i=60^{\circ}$ under illumination of an incident wave with (a) magnetic field on $z$-axis produce total retroreflection with different phases. For illustrative purposes, the resulting magnetic field for reflected waves with (b) $\varphi_{\rm r}={0}^{\circ}$, (c) $\varphi_{\rm r}={90}^{\circ}$, and (d) $\varphi_{\rm r}={180}^{\circ}$ are shown.}
\label{comsol_fld}
\end{figure}

To verify Eq.~\r{sandwich_eqv}, we use a Comsol model (\mbox{Fig.~\ref{eqv_circ}}) with the following parameters: \mbox{$t_{1}=t_{2}=t_{3}=0.5$ mm}, $\varepsilon_{1}=\varepsilon_{2}=2.7$, $\varepsilon_{3}=5.5$, $\theta_{\mathrm{i}}=60^{\circ}$, $f=29$~GHz. The results of the simulation show that the incident [\mbox{Fig.~\ref{comsol_fld}(a)}] and reflected [\mbox{Figs.~\ref{comsol_fld}(b)--(d)}] magnetic fields propagate in the opposite directions, as desired. It is also important to mention tunability of the reflection phase, as it is illustrated in \mbox{Figs.~\ref{comsol_fld}(b)--(d)}. This parameter will take an important role in the design of the metasurface for surface waves decoupling.

\subsection{Discretization tolerance}
\begin{figure}[!h]
\centerline{\includegraphics[width=0.9\columnwidth]{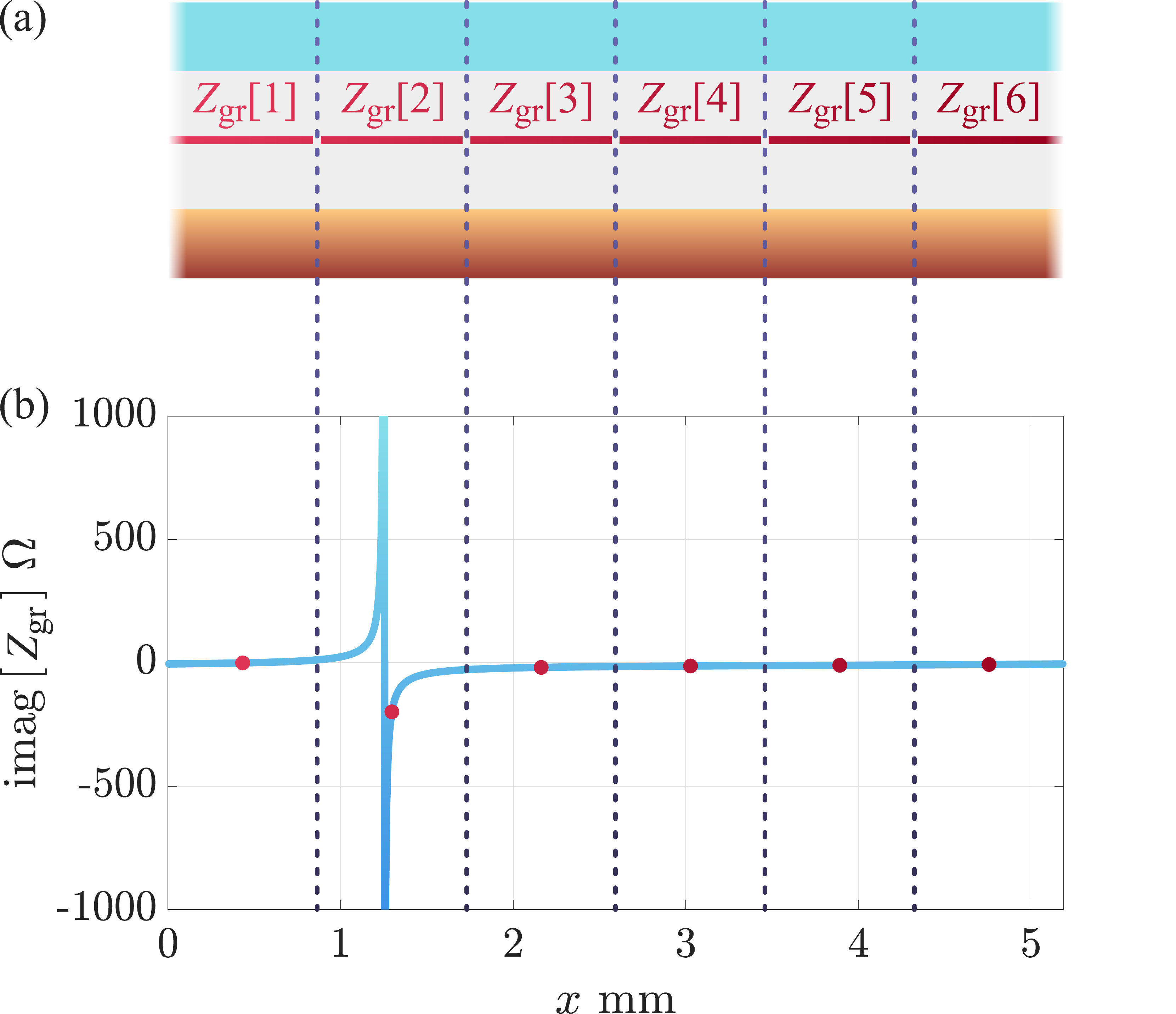}}
\caption{For a practical realization, the metasurface should have a discrete grid impedance profile. This is achieved by (a) splitting the surface evenly and (b) assigned each section a constant surface impedance which is sampled from the continuous surface impedance profile.}
\label{descr_imp}
\end{figure}

In the previous subsection, we considered a continuously varying grid impedance [see the blue curve in \mbox{Fig.~\ref{descr_imp}(b)}], which is not possible to implement in practice, because unit cells have finite sizes. To solve this issue, the grid impedance should be discretized to a finite set of points \cite{Wang_2018_printable}. An example of a discretized impedance is shown in \mbox{Fig.~\ref{descr_imp}(b)} (orange points). 

\begin{figure}[!h]
\centerline{\includegraphics[width=0.9\columnwidth]{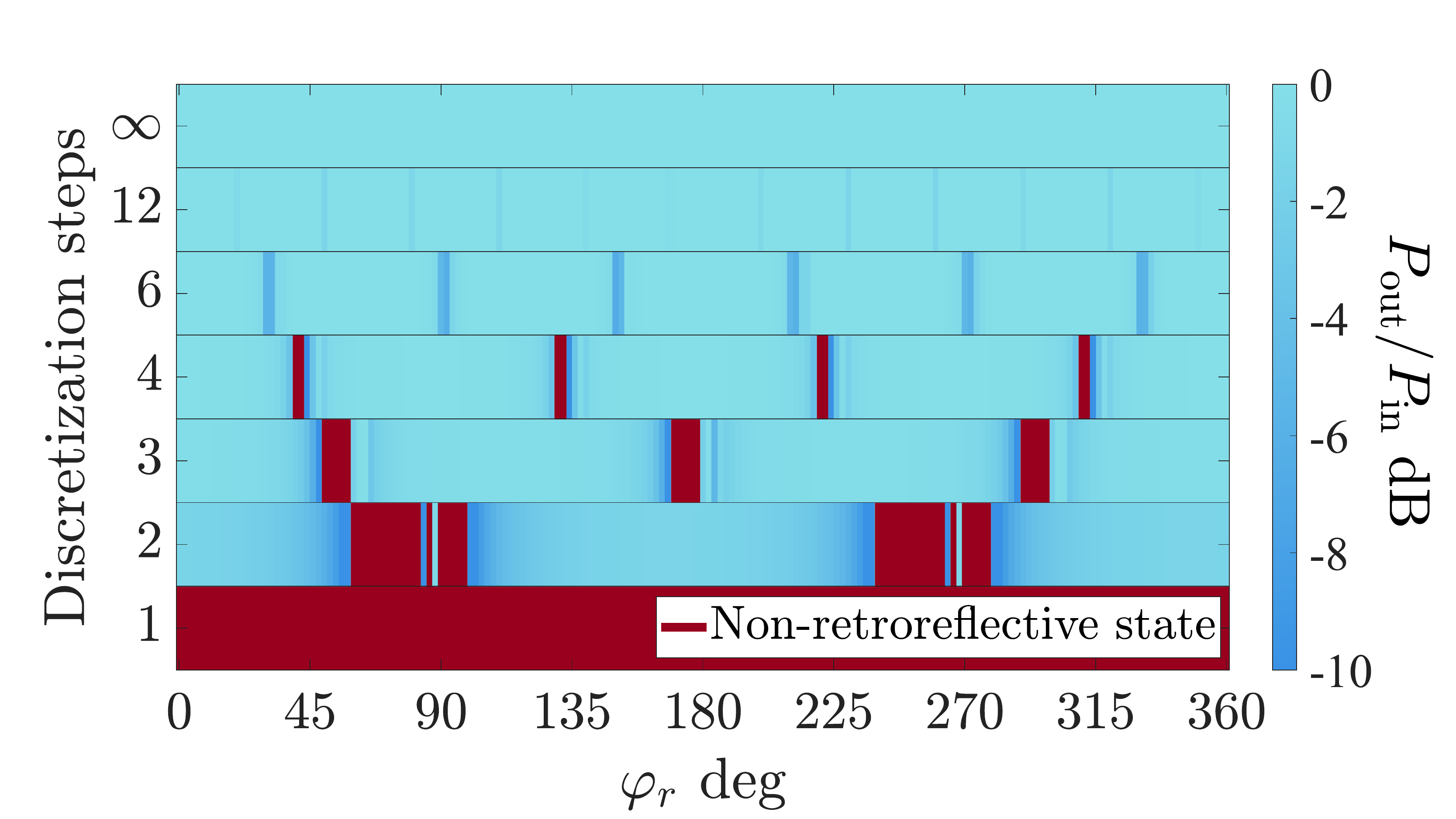}}
\caption{Effect of the number of discretization steps on the power reflected for the incidence angle $\theta_i=60^{\circ}$ with different reflection phases $\varphi_{\rm r}$. Non-retroreflective states, where the metasurface produces specular reflection, appear for small numbers of discretization steps.}
\label{discret_efficiency}
\end{figure}

The performance of the discretized metasurface will be affected by the number of discretization steps, as shown in \mbox{Fig.~\ref{discret_efficiency}}. As expected, it is not possible to achieve retroreflection using a single discretization step, as the resulting structure does not compensate the incidence wave phase, and it behaves as a conventional specular reflector. However, as the number of discretization steps increases, the metasurface can be tuned to achieve retroreflection with acceptable performance. Based on the information provided in \mbox{Fig.~\ref{discret_efficiency}}, using six discretization steps offers a good trade-off between performance for almost every value of $\varphi_{\rm r}$ (required for fine tuning) and number of required different impedance values.

\section{Metasurface implementation}

\begin{figure}[!h]
\centerline{\includegraphics[width=0.9\columnwidth]{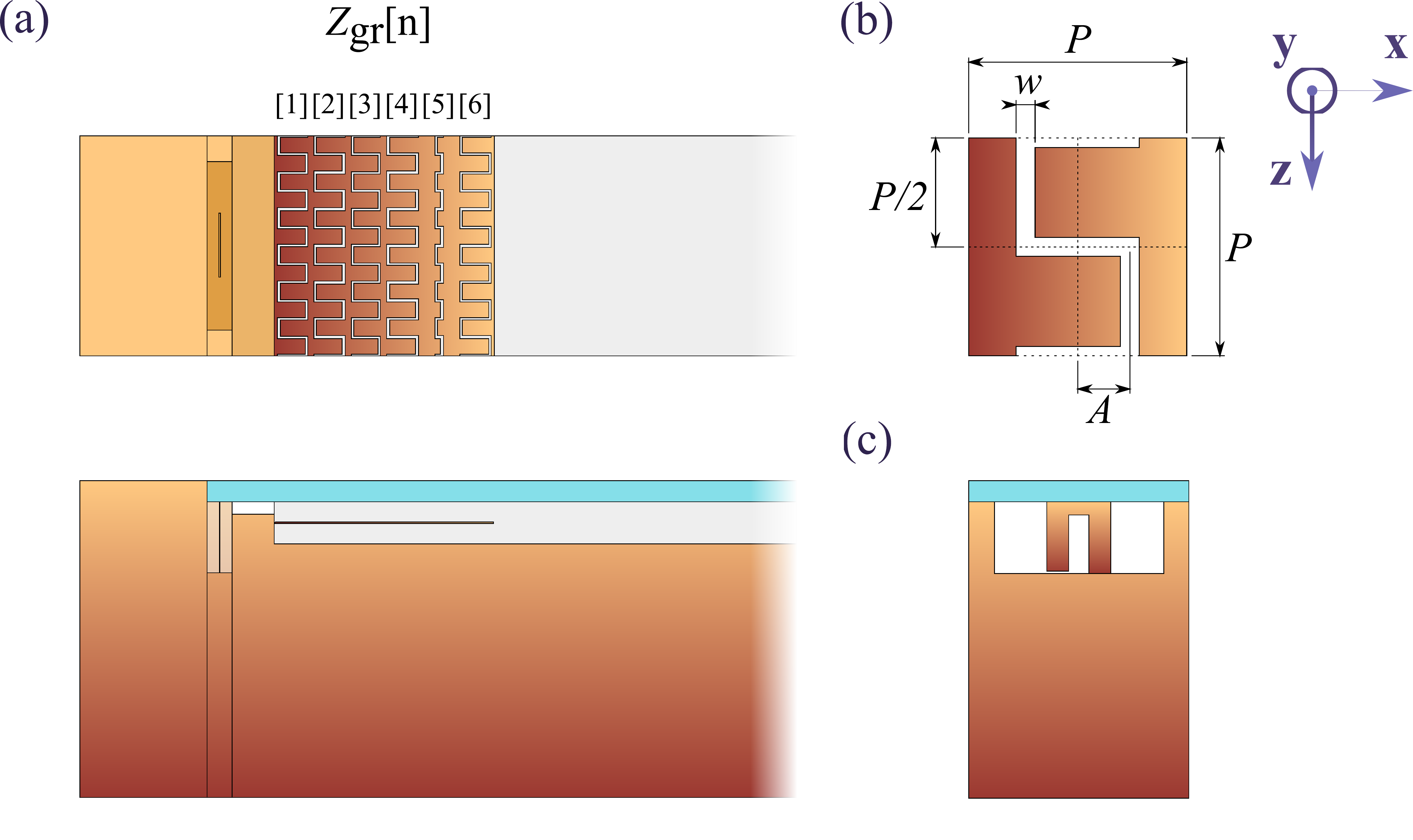}}
\caption{(a) As a proof of concept, a model periodic along the $z$-axis (with the period $D=5.19$~mm) is used to study  the performance of the proposed structure discretized  into six sub-cells. The total length of the model is about 154~mm, representing the region close to the antenna. (b) The metasurface is discretized based on a meandered-slot topology, whose unit-cell geometry is defined by the unit cell period $P$, the gap width $w$, and the meander amplitude  $A$. (c) As an example, the proposed structure is designed to improve the radiative properties of a folded dipole antenna located below the glass layer in an air-filled gap near the dielectric layer, and partially bounded by PEC walls.}
\label{periodic_structure}
\end{figure}

\begin{figure}[!h]
\centerline{\includegraphics[width=1\columnwidth]{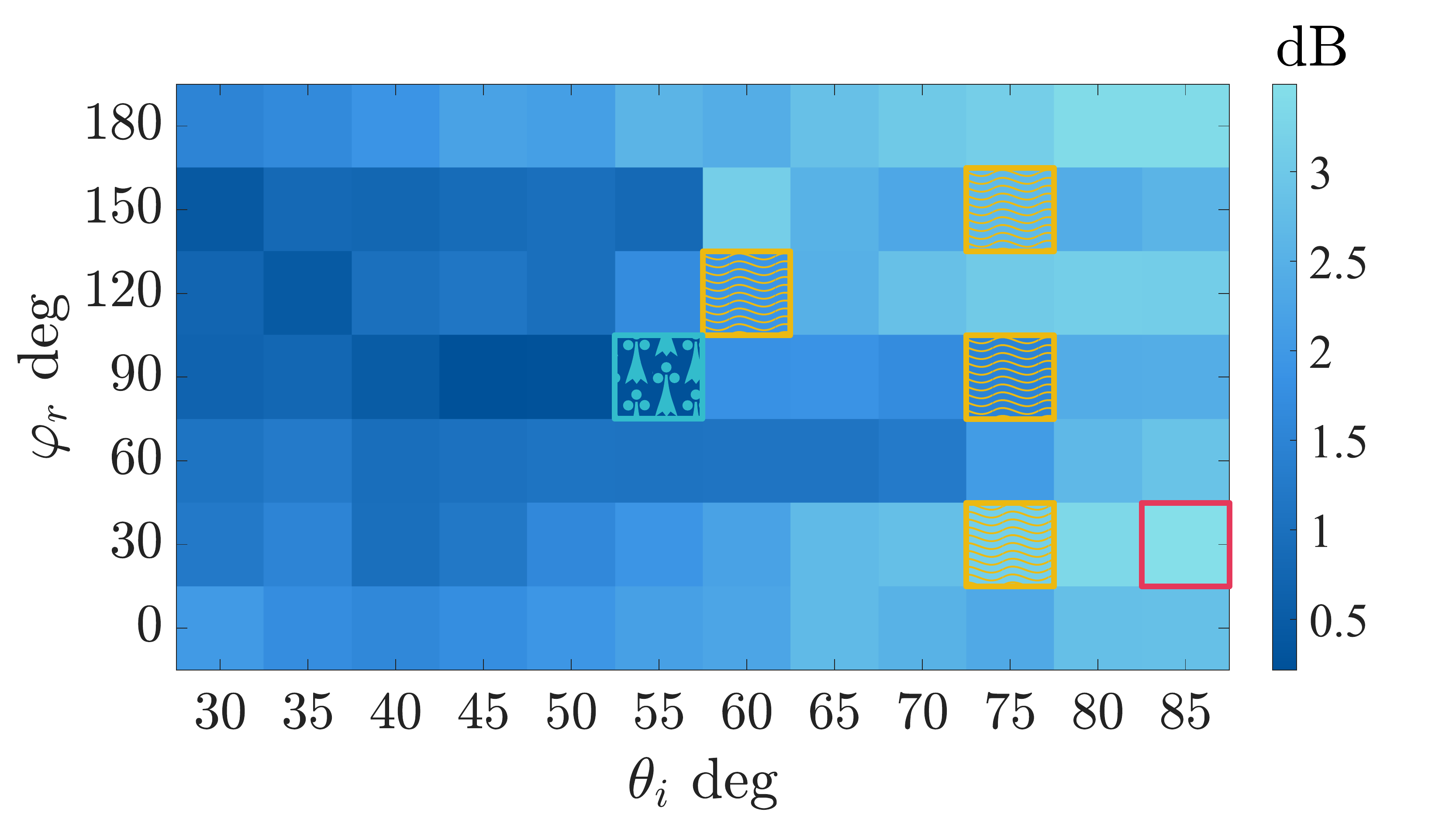}}
\caption{Average improvement for the region of interest from $-45^{\circ}$ to $45^{\circ}$. The best performance is achieved using \mbox{$\theta_i=85^{\circ}$} and \mbox{$\varphi_r=30^{\circ}$}, denoted as a square, while the worst scenario is found with \mbox{$\theta_i=55^{\circ}$} and \mbox{$\varphi_r=90^{\circ}$} (Ermine pattern). Due to numerical stability issues, wave-pattern  results were estimated as a mean value from shifting $\varphi$ about $\pm5^{\circ}$ while preserving $\theta_i$.}
\label{angle_optimization}
\end{figure}

The performance of the proposed retroreflective sheet is evaluated in the structure shown in \mbox{Fig.~\ref{periodic_structure}}, where a folded dipole excites parasitic surface waves along the dielectric and glass layers. The retroreflective metasurface is located in the middle of the dielectric layer, aligned with a metallic wall that prevents wave propagation below the metasurface. Using six discretization sub-cells, the optimal values for $\theta_i$ and $\varphi_r$ were found using numerical (CST) simulations. The used optimization criterion was the maximized radiation in the $y$-direction with the beamwidth of $\pm 45^{\circ}$, compared to the performance in the same range without the retroreflector. The retroreflective layer was tested for different values of $\theta_i$ in the range between $30^{\circ}$ and $85^{\circ}$ using $5^{\circ}$ steps; in combination with values of $\varphi_r$ in the range between $0^{\circ}$ and $180^{\circ}$ with $30^{\circ}$ steps. The simulation results for each combination of $\theta_i$ and $\varphi_r$ are  collected in \mbox{Fig.~\ref{angle_optimization}}. Due to numerical singularities, some parameter combinations could not be tested reliably, and we display the averaged values using $\varphi_r\pm5^{\circ}$ instead. Nevertheless, these estimated values are far away from the optimal region of interest, close to  $\theta_i=85^{\circ}$. As a result, the optimal performance using ideal discretized surface impedances was found with $\theta_i=85^{\circ}$ and $\varphi_r=30^{\circ}$.

Due to its capacitive nature, the retroreflective metasurface can be realized using the meandered-slot topology \cite{Wang_2018_asymmetry}, illustrated in \mbox{Fig.~\ref{periodic_structure}(b)}. While the period of the meandered-slot unit cell $P$ is fixed to be one sixth of the grid impedance period $D$, the other  parameters (the gap width $w$ and the  meander amplitude $A$) had to be tuned for every sub-cell to achieve the required reactance. The designed values are collected in Table~\ref{tab:cell_dimensions}.


\begin{table}[!h]
    \centering
    \begin{tabular}{|c|c|c|c|c|c|c|}
        \hline
         Subcell & 1 & 2 & 3 & 4 & 5 & 6  \\
         \hline
         ${\rm Im} \left[ Z\@{gr} \right]$ $\Omega$ & -47 & -254 & -68 & -62 & -58 & -55  \\
         ${\rm Im} \left[ Z\@{s} \right]$ $\Omega$  & 60 & -1255 & -54 & -18 & 0.86 &   20 \\
         $A$ mm & 0.34 & 0.26 & 0.34 & 0.32 & 0.34 & 0.33 \\
         $w$ mm & 0.06 & 0.065 & 0.065 & 0.055 & 0.063 & 0.058 \\
        \hline    
          
    \end{tabular}
    \caption{Parameters of the designed  retroreflector sub-cells based on the  meandered-slot topology of \mbox{Fig.~\ref{periodic_structure}(b)}.}
    \label{tab:cell_dimensions}
\end{table}

\section{Results}

\begin{figure}[!h]
\centerline{\includegraphics[width=0.6\columnwidth]{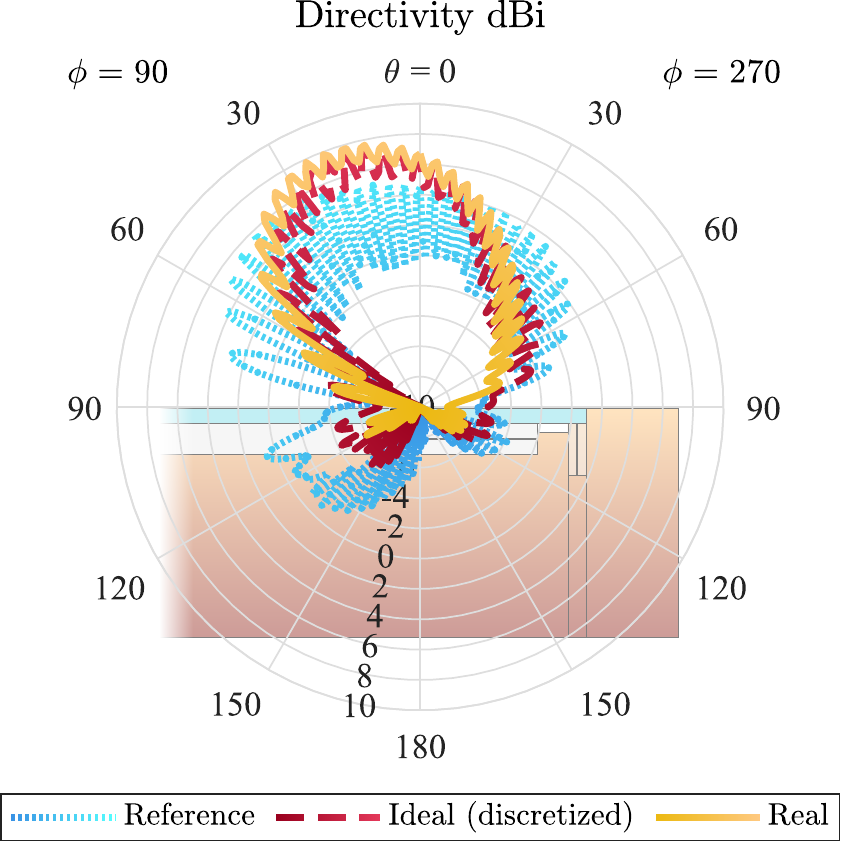}}
\caption{The use of a retroreflective surface in a multilayer structure improves the directivity in the angular range  $\left|\theta \right|\leq 45^{\circ}$. An optimized implementation based on meandered slots improves the performance of the ideal discretized impedance sheets.}
\label{directivity_comparison}
\end{figure}

The efficiency of the proposed solution can be estimated by comparing the device shown in  \mbox{Fig.~\ref{periodic_structure}} with and without the retroreflective metasurface. \mbox{Figure~\ref{directivity_comparison}} contrasts the far-field directivity of the reference antenna, that with the ideal discretized metasurface, and with its meandered-slot implementation at the reference frequency of 29~GHz. The use of the retroreflective sheet decreases the radiation along the multilayered structure plane, while the directivity is improved in the direction normal to the layers, as desired. Small optimizations of each sub-cell parameter allows us to maintain and improve the resulting radiation pattern. 

\begin{figure}[!h]
\centerline{\includegraphics[width=0.9\columnwidth]{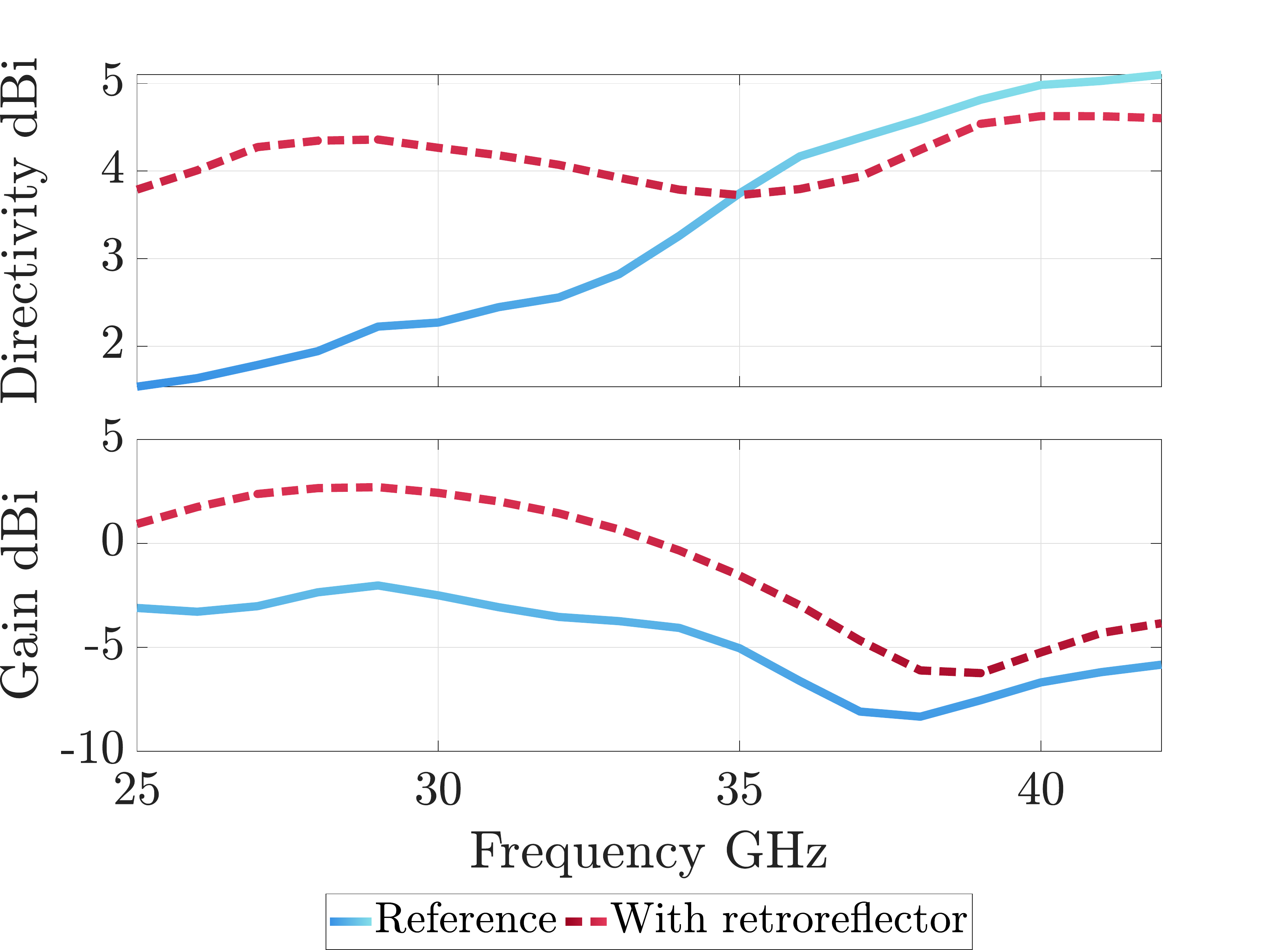}}
\caption{As a result of placing the metasurface inside the dielectric layer, the average directivity in the range $\theta_i<45^{\circ}$ stabilizes around 4 dBi in the frequency range between 25 and 42 GHz. Similarly, the antenna gain is improved by the use of the retroreflective sheet.}
\label{directivity_frequency}
\end{figure}

As a complement, \mbox{Fig.~\ref{directivity_frequency}} offers a deeper insight on how the retroreflective layer performs across the frequency range of 25--42 GHz. The capacitive, non-resonant nature of the retroreflector grants a broadband frequency response in terms of the average directivity (around 4~dBi), outperforming the out-of-the-box device for frequencies lower than 35~GHz, while for higher frequencies the performance is similar. In terms of gain, \mbox{Fig.~\ref{directivity_frequency}} proves that a retroreflective layer can improve the performance of a 5G device, with a gain improvement oscillating between 1.3 and 5.4~dB.

\begin{figure}[!h]
\centerline{\includegraphics[width=0.9\columnwidth]{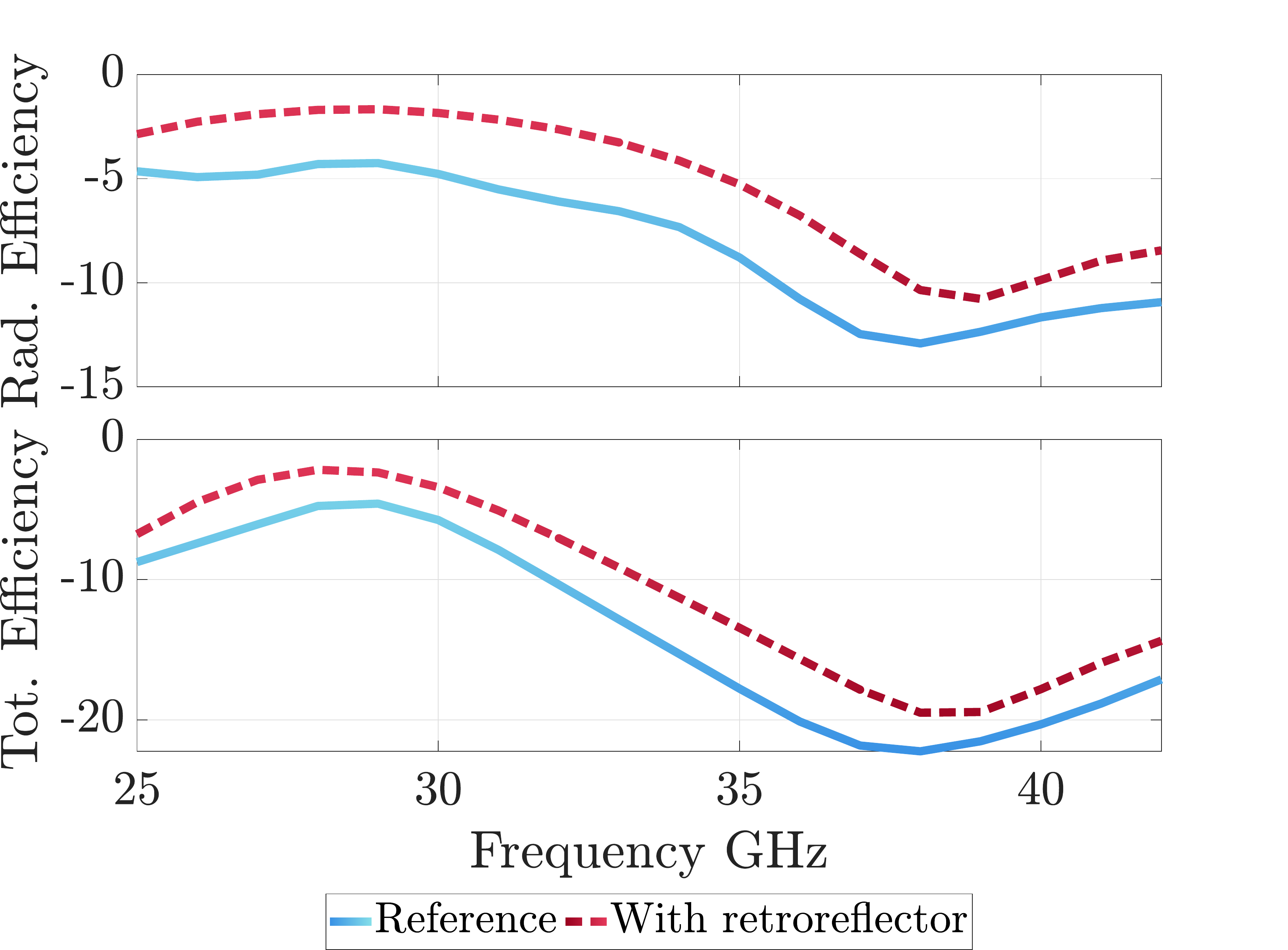}}
\caption{The use of the proposed metasurface allows us to increase the radiation efficiency (power radiated in  free space without antenna impedance matching, on dB scale) around 2~dB compared to the reference structure. With a similar trend, the retroreflective sheet improves the total efficiency (also considering antenna matching, in dB scale).}
\label{efficiency_frequency}
\end{figure}

In terms of radiated power, \mbox{Fig.~\ref{efficiency_frequency}} shows that the combined antenna-retroreflector systems are capable to radiate power more efficiently. The radiation efficiency, without considering antenna impedance matching, increases by 1.57--4.01~dB by using a retroreflector. Similar results can be observed for the total efficiency, including impedance matching, where the retroreflector improves the performance by 1.99--4.47~dB.

\section{Conclusions}

In this work we have introduced a retroreflective surface capable to improve the radiation pattern and efficiency of  smartphone antennas in the mmWave frequency range. By using the concept of retroreflection, a metasurface was designed to create a virtual retroreflective surface over the glass surface, with adjustable reflection phase. The resulting grid impedance matches the TM-polarized wave inside the dielectric layer, reducing the parasitic wave propagating along the smartphone body. The retroreflecting element prevents parasitic channeling of the antenna energy into surface waves in and behind the glass layer and directs the radiation into the desired direction, in contrast to   conventional periodical stop-band structures that only forbid propagation of surface waves.

The discretization of the impedance profile, required for a plausible realization, was realized as a trade-off between performance and complexity with a sweet spot of six sub-cells. The phase of the retroreflected wave became a relevant parameter for fine tuning and for choosing a favourable set of grid impedance values. A topology based on meandered slot unit-cells was selected due to the pure capacitive nature of the required grid impedance. 

To achieve a compact design, the resulting metasurface only covers one period of the surface impedance, without compromising  performance.  The proposed structure shows improvements in radiative and efficiency terms, granting  improvements in antenna directivity and gain around 2~dB and 5~dB, respectively; while the antenna efficiency increases around 4~dB. The non-resonant nature of the metasurface allows improvement of the antenna performance in a wide frequency range.

\bibliographystyle{IEEEtran}
\bibliography{reference}




\end{document}